\documentclass[12pt]{article}
\usepackage{epsfig}

\def\Title#1{\begin{center} {\Large {\bf #1} } \end{center}}

\begin{document}

\Title{Quantum Vacuum Structure and Cosmology}

\bigskip\bigskip


\begin{raggedright}  

{\it Johann Rafelski\index{Rafelski, J.}, Lance Labun\index{Labun, L.}, Yaron Hadad\index{Hadad, Y.},\\
Departments of Physics and Mathematics, The University of Arizona\\ 85721 Tucson, AZ, USA, and\\
Department f\"ur Physik, Ludwig-Maximillians-Universit\"at M\"unchen \\ 
Am Coulombwall 1, 85748 Garching, Germany

\vspace*{0.5cm}
Pisin Chen\index{Chen, P.},\\
Leung Center for Cosmology and Particle Astrophysics and\\
Graduate Institute of Astrophysics and\\
Department of Physics, National Taiwan University, Taipei, Taiwan 10617, and \\
Kavli Institute for Particle Astrophysics and Cosmology, SLAC National Accelerator Laboratory, Stanford University, Stanford, CA 94305, U.S.A.
}
\bigskip
\bigskip
\end{raggedright}

{\bf Introductory Remarks}\\
Contemporary physics faces three great riddles that lie at the intersection
of quantum theory, particle physics and cosmology.  They are
\begin{enumerate}
\item The expansion of the universe is accelerating -- an extra factor of two appears in the size.
\item Zero-point fluctuations do not gravitate -- a matter of 120 orders of magnitude
\item The ``True'' quantum vacuum state does not gravitate.
\end{enumerate}
The latter two are explicitly problems related to the interpretation and the 
physical role and relation of the quantum vacuum with and in general relativity.
Their resolution may require a major advance in our formulation and 
understanding of a common unified approach to quantum physics and gravity.  
To achieve this goal we must develop an experimental basis and much of 
the discussion we present is devoted to this task. 

In the following, we examine the observations and the theory
contributing to the current framework comprising these riddles.
We consider an interpretation of the first riddle within 
the context of the universe's quantum vacuum state, and propose 
an experimental concept to probe the vacuum state of the universe.

\vspace*{0.5cm}
{\bf  The Riddles}\\
\emph{\bf Riddle 1.}
Data indicate that the universe expansion is accelerating and comprehensive
studies of cosmological and astrophysical observables determine the object
driving the expansion to have an equation of state~\cite{Frieman,Komatsu}
$$w\equiv p/\rho = -0.94 \pm 0.1,$$  
incompatible with 
normal matter which has $w\le 1/3$ --- hence the term dark energy, distinct 
from dark matter. Many works offer an explanation  introducing new types of 
dynamical fields in order to provide the requisite behavior of the dark 
energy density (for review see~\cite{Copeland:2006wr})
or even both dark energy and dark matter at the same time~\cite{Bilic:2001cg}.  
However, theories entailing dynamics in the evolution of dark energy have been
severely constrained~\cite{Serra:2009yp}.  Due to the 
homogeneity of its distribution in space and time, the dark energy is most 
consistent with a cosmological constant $\Lambda$ which enters the 
Einstein equation 
(gravitational metric $-,+,+,+$, 
signs following convention of Weinberg~\cite{Weinberg72})
\begin{equation}\label{EEqs1}
R_{\mu\nu}-\frac{1}{2}g_{\mu\nu} R+g_{\mu\nu}\Lambda = -8\pi G T_{\mu\nu}.
\end{equation}
Considering the value 
\begin{equation}\label{dark}
\lambda := \frac{\Lambda}{8\pi G}\simeq (2.6\pm 0.6\,{\rm meV})^4 
	\simeq 3.4\pm 0.4 \times 10^{-10}\,{\rm J/m}^3,
\end{equation}
dark energy  amounts to a strikingly small 
energy density $g^{00}\lambda$, corresponding to about 2 protons per m$^3$, 
or the energy content in the electric field of magnitude 8.3\,V/m, and yet,
given the homogeneous distribution it is the dominant energy content in present 
day expansion diluted Universe. This dominance is a relatively recent phenomenon, 
for most of its   history (on logarithmic  time scale) the Universe has
been dominated by matter, and earlier on, radiation, a negligible 
component today.

Moving on, general relativity and quantum field theory have not yet been 
made consistent, and the most striking symptoms of this situation are the 
two other riddles.

\vspace*{0.5cm}
\emph{\bf Riddle 2.}
The vacuum fluctuations of the known matter fields cannot gravitate:
a simple summation of the zero-point energy of e.g. electron-positron Dirac 
field shows
\begin{equation}\label{zeropt}
\langle \epsilon\rangle_{\rm matter} = 
-2_s\cdot 2_p\cdot\int_0^{M_{\rm Pl}} dk \frac{4\pi k^2}{(2\pi)^3}
	\frac{1}{2}\sqrt{k^2+m^2} \simeq -\frac{M_{\rm Pl}^4}{16\pi^2},
\end{equation}
which using the Planck mass $M_{\rm Pl} = 1.2\times 10^{19}$\,GeV gives 
$$ \langle \epsilon\rangle_{\rm matter} \simeq 10^{120}\lambda, \qquad
-8\pi G \langle \epsilon\rangle_{\rm matter} = \frac{c}{\hbar}\frac{4}{8\pi G}\,
. $$
No known framework, realistic formulations of super-symmetry included, cancels
near to 120 orders of magnitude.  Similarly, it is hard to imagine how to 
cancel a $1/G$ term in the curvature of the Universe. Consistency 
between the  present day  quantum field theory, and gravity, appears to be 
impossible.

\vspace*{0.5cm}
\emph{\bf Riddle 3.}  Thus for whatever reason the True Vacuum does not 
gravitate.
The vacuum we observe in the universe---whether the True Vacuum or a 
False (quasi-stable) Vacuum---is highly structured by 
electric, weak and color charged interactions. Recall that from electroweak 
(EW) theory and quantum chromodynamics (QCD) we derive many physical 
properties and structures this vacuum must have.
Let us consider some examples:
\begin{enumerate}
\item
In QCD, color confinement requires its vacuum state be defined by
\begin{equation}
\langle F_{\mu\nu}^a\rangle_V \equiv 0, \quad\mathrm{i.e.}~~ 
\langle \vec E^a \rangle_V \equiv 0~~\mathrm{and}~~\langle \vec B^a \rangle_V \equiv 0.
\end{equation}
Yet evaluating the glue condensate, we have (color traces implied)
\begin{equation}\label{glue-cond}
\langle\frac{\alpha_S}{\pi}G^2\rangle_V = \left[ 330(50)\,{\rm MeV}\right]^4,
\end{equation}
showing that the confining vacuum must be dominated by color-magnetic
fluctuations
\begin{equation}
\langle B^2 \rangle = \frac{1}{2}\langle G^2 \rangle + \langle E^2\rangle.
\end{equation}
\item
  The fluctuating color fields will induce a quark 
condensate and indeed one finds
\begin{equation}
\langle \overline{u}u + \overline{d}d\rangle_V = 
	-2\left[225(25)\,{\rm MeV}\right]^3
\end{equation}
\item
 The spontaneous symmetry breaking  structure of the EW vacuum implies an 
omni-present condensate
\begin{equation}
\langle H \rangle = (\sqrt{ 2} G_F)^{-1/2}\simeq 0.2462\,\mathrm{TeV}.
\end{equation}
We remark that the coupling of the Higgs to the top quark is almost unity: $g_t = 0.99$
so that $M_t = g_t{\langle H\rangle}/{\sqrt{2}}=0.1724 $\,TeV
indicating a relationship between the QCD and EW vacuum structure, yet to be discovered.
\item
Amidst the widely different energy scale the final puzzle piece is neutrino 
mass difference in the range $\Delta m_{\nu} = 10 - 100~\mathrm{meV}.$  
How neutrinos acquire this mass is hotly debated. Since all other particle 
masses are properties of the vacuum structure, it is natural to assume that 
also neutrino masses are. This suggests existence of vacuum structure beyond 
the Standard Model with a scale corresponding to that of dark energy.
\end{enumerate}


\vspace*{0.5cm}
\emph{\bf The True and The False Vacuum}\\
A natural place in the present theoretical paradigm to
`find' a cosmological constant is in the 
quantum vacuum~\cite{Weinberg:1988cp}.
The dark energy in the form of Einstein's cosmological 
constant stands next to an energy momentum tensor $T_{\mu\nu}$,
and should be inherent to the vacuum expectation values of the energy 
momentum tensor.
Any energy momentum tensor which is not traceless, for which 
the energy density $T^{00}>0$, and the pressure 
components $T^{ii}<0, i=1,2,3$ i.e. have wrong, non-matter-like sign, 
can be decomposed to make the inherent
dark energy component visible. This is done by separating 
the energy-momentum trace
\begin{equation}\label{trsep}
T_{\mu\nu}=\widetilde T_{\mu\nu}+g_{\mu\nu}\frac{T_{\alpha}^{\alpha}}{4},
\end{equation}
and the Einstein equation is presented in the form
\begin{equation}\label{tr+EEqs1}
\frac{1}{8\pi G}\left(R_{\mu\nu}-\frac{1}{2}g_{\mu\nu} R\right) = 
-\widetilde T_{\mu\nu}-g_{\mu\nu}
    \left(\lambda +\frac{T_{\alpha}^{\alpha} }{4}\right).
\end{equation}
Such an effective dark energy term naturally arises in  nonlinear 
electrodynamics~\cite{Labun:2008qq}.

The measurement of a finite value of dark energy tells us that 
the quantum ground state of the present Universe  
is {\bf not} the true lowest energy state. We interpret 
the cosmological constant to be the difference in the energy density 
of the present state universe $|\Psi\rangle$  with regard to  the 
True Vacuum $|{\rm TV}\rangle$:
\begin{equation}
\lambda = 
\langle\Psi|\mathcal{H}|\Psi\rangle - \langle{\rm TV}|\mathcal{H}|{\rm TV}\rangle.
\end{equation}
The Hamiltonian $\mathcal{H}$ is that of the `complete theory,' whose 
non-vanishing expectation value indicates that the 
universe is trapped by a large potential barrier in 
the (slightly) excited state $|\Psi\rangle$.
This finding  does not contradict our usual assumption that the 
vacuum energy is zero since up to the effect of gravity, 
the measurement of energy density of the vacuum is shift-invariant.
That is, when we study properties of the vacuum and 
the laws of physics, we can assume that a quasi-stable quantum ground 
state has a vanishing energy density.

\vspace*{0.5cm}
\emph{\bf Recovering the True Vacuum}\\
Under the hypothesis that the dark energy is a consequence of the universe
being trapped in a false vacuum state, we must discuss the consequences of 
decaying to the True Vacuum.  Such ideas have been widely 
discussed before~\cite{Stone:1975bd,Coleman:1977py,Callan:1977pt,Linde:1981zj}. 
We note in analogy that the presence of an electric field in the vacuum 
renders it unstable to pair production at a rate that becomes extremely rapid 
as the Schwinger field $E_0=m^2/e$ is approached.

Were we to succeed in inducing a transition to the True Vacuum
in some finite volume element, it is likely in our present situation not to
be catastrophic: the amount of energy released during local vacuum decay 
is too small to maintain combustion, because the apparent stability of the
vacuum suggests that the rate of the decay is small 
and therefore the height (and the width of 
the barrier separating the false and the True vacuum states) can be
much higher than the energy gained in the process,
$$  \Delta V_{\rm barrier} \gg \lambda. $$ 
Thus to maintain combustion we need to 
keep the environment that catalyzes the decay. In other words we must arrange 
an experimental environment in which the dark energy can burn.

\emph{\bf Frames of Reference and Mach's Principle}\\
The question is posed, in what frame is the energy recovered which is 
freed should the false  vacuum decay be observed?
As remarked during the discussion session of this lecture by~\cite{BMTC}, the
vacuum state in a Lorentz-covariant theory cannot be a preferred frame 
of reference.  Therefore, we cannot detect motion of any apparatus
with respect to the vacuum, and the concept of a relative observer-vacuum
velocity is ill-defined, as is the production of energy in vacuum combustion.
 
The difficulty in conception here is encapsulated in Mach's Principle,
which as formulated by Einstein is the statement that
\begin{equation}\label{MP}
\mathrm{Mach's~Principle:~Inertia~is~measured~relative~to~the~fixed~stars.}
\end{equation}
Contemporary cosmology assumes Mach's Principle in praxis, 
by \emph{identifying} the frame of reference of the cosmic microwave 
background (CMB) as `the frame of the universe.'  
This identification is defensible considering that 
the existence of a very high temperature (symmetry-restoring) 
heat bath early in the universe provides a preferred frame for the universe, 
which is the frame at rest with respect to the heat bath. The 
remnant photons in the CMB have with great likelihood 
preserved an observable link to this preferred frame.

While we cannot 
define a position in the sense of an absolute frame, we can not only
measure our velocity with respect to cosmic frame but 
we can and indeed must presume that 
the global false vacuum state is defined with respect to this cosmological 
frame of reference. 
Dark energy is experimentally determined in this cosmological sense. 
Therefore, the dark energy, interpreted as energy of 
the false vacuum, would be released in this same frame, and the relative
motion of the experiment with respect to the CMB is the correct relationship
to consider.

\vspace*{0.5cm}
\emph{\bf Experimental observable}\\
We imagine an experimental device that induces the transition to the 
True Vacuum and is crossing the universe.  The energy of the false vacuum 
quench should in some part be converted to radiation.  If we assume that all 
energy turns into radiation, we would have a temperature 
\begin{equation}
\lambda = \frac{E}{V}=\frac{\pi^2}{30}2_sT_{\rm eq}^4, \quad 
T_{\rm eq}=0.9\times 2.6~\mathrm{meV/k_B} = 23\pm 6\,\mathrm{K},
\end{equation}
well above that of the cosmic background radiation.
Launched from Earth, the device travels at high speed $v\simeq 300$ km/s 
with respect to the cosmic microwave background.  Inducing vacuum decay in 
the experimental volume, the device observes a heat flux:
\begin{equation}
J_Q = \lambda v =  10^{-4} \frac{\rm W}{\mathrm{m}^2} 
\end{equation}

\vspace*{0.5cm}
\emph{\bf Vacuum quench by QCD matter}\\
Transition to the True Vacuum could be induced by providing energy in the form
of an applied field, which deforms the effective potential and may reduce the
barrier between states.

The large energy density of the QCD vacuum 
(recall Eq.\,(\ref{glue-cond})) suggests we study
the volume transited by atomic nuclei to determine if it is converted to the 
True Vacuum.  If so, a large material object could quench the false vacuum, in
which case the experimental signature is that the object cannot drop to 
arbitrarily low temperature. A suitable environment for such an experiment is
extraterrestrial space,  so as  to get 
out from under the 1000 g/cm$^2$ of air necessarily 
between any ground-based experiment and the false vacuum.

The dark side of the moon does  not receive sunlight for 14 days at a time, 
and remote sensing  could be setup to detect the appropriate 
temperature periodicity, which is a combination of the varying direction of 
the moon's motion with respect to cosmological frame and heating during 
solar days.  The minimum moon temperature $T = 15-40$K 
in deep craters near the dark polar regions is notable in this regard.

A man-made experiment could involve the launching into space a long metal rod.
The end of the rod oriented in the direction of motion relative to the cosmic
frame of reference of the universe should experience a higher temperature, 
being the site of active vacuum combustion. Note that if successful such an 
experiment would resolve many cosmological questions.

\vspace*{0.5cm}
\emph{\bf Vacuum quench by Electro-Weak Interactions}\\
The vacuum of quantum electrodynamics (QED) is not thought to have a rich 
structure.  However the connection between QED and weak interactions is very 
complex and generates vacuum properties which have not yet been understood 
well even though they stand at the origin of the unification 
of QED and weak interactions into electro-weak theory. 
The similarity of the dark energy and neutrino mass scale, both beyond the 
standard model, inspires many to consider that the dark energy originates in 
the electro-weak sector.

A potential tool to probe neutrino related defects in the electro-weak vacuum
 structure could be an ultra intense pulsed laser. The
natural wavelength of light is comparable to the Compton wavelength
\begin{equation}
l_\nu = 2\pi \frac{\hbar c}{m_{\nu}}= 1~\mu\mathrm{m}
\quad \mathrm{for}\quad m_{\nu} = 1.24~\mathrm{eV}.
\end{equation}
Strong electromagnetic fields generated by extreme pulsed lasers 
with wavelengths at the micron appear to be natural 
tools capable to quench the false  neutrino  vacuum.
However, even if this were happening today we are not quite able to 
detect the subtle effects.  An exception to this arises should the 
weak interaction stability properties be modified. In fact, there have been 
sporadic and unconfirmed reports that strong laser fields can modify 
electro-weak decays and this subject warrants some attention.

\vspace*{0.5cm}
\emph{\bf In lieu of conclusions: Back to the \ae ther}\\
Over the past 50 years of continuous development of the standard model and
improved understanding of the quantum vacuum structure we have in essence
made a full circle: we are at the verge of recognizing that the Universe
is in essence filled with an \ae ther, which respects locally the 
Lorentz symmetry and thus can play a role only when we carve out a ponderable
domain where modifications occur, or consider the entire Universe we can see.
 
It is not generally known how Einstein changed in his time his views: 
it is widely reported that he rejected \ae ther as unobservable when formulating 
special relativity. Within 15 years, once the introduction
of general relativity and cosmology was achieved,
he writes in a 1920 letter to Lorentz~\cite{Einstein-lett}
\begin{quote}
It would have been more correct if I had limited myself, in my earlier
publications, to emphasizing only the non-existence of an \ae ther
velocity, instead of arguing the total non-existence of the \ae ther, 
for... I can see that with the word \ae ther we say nothing else than 
that \emph{space has to be\\ viewed as a carrier of physical qualities.}
\end{quote}
[our emphasis] and again a few months later in his review  
prepared for presentation in front of Lorentz in Leiden, he discusses
in depth the \ae ther and he closes:~\cite{Einstein-lect}
\begin{quote}
...\ae ther may not be thought of as endowed with the quality 
characteristic of ponderable media, as consisting of parts which may 
be tracked through time.  The idea of motion may not be applied to it.
\end{quote}
Today we effectively accept these Einstein's views, and yet the last
point, put forward within a fully classical framework, has turned out to be 
untenable once \ae ther was identified implicitly as the quantum vacuum.  
Almost everybody argues that  
the ground state, the vacuum, can be viewed as a ponderable medium. 
We speak of melted vacuum, and formation of quark-gluon plasma. We 
modify the vacuum fluctuations and measure quasi-force named after Casimir,
and we apply strong EM fields to induce vacuum decay. In these examples, 
the vacuum is locally defined, and only if this is true can we expect to 
to detect local vacuum changes.  
However, the presence of quantum physics discovered after 
Einstein's reintroduction of the \ae ther and discussion of its 
properties are of substance because only in this way can we retain
quasi-stability of local parts of the vacuum.

We hope that the present discussion will encourage work towards discovery of 
the false (dark energy) vacuum decay.

\bigskip
{\bf Acknowledgments} JR, LL and YH are supported by a grant from: the
U.S. Department of Energy  DE-FG02-04ER41318, and by the 
DFG Cluster of Excellence  MAP (Munich Centre of Advanced Photonics); 
PC is supported by US DOE (Contract No. DE-AC03-76SF00515) and 
the National Science Council of Taiwan (Grant
No. 95-2119-M-009-026).


\begin{thebibliography}{99}

\bibitem{Frieman}
  J.~Frieman, M.~Turner, and D.~Huterer,
  Ann.\ Rev.\ Astron.\ Astrophys.\ {\bf 46}, 385 (2008).
  
\bibitem{Komatsu}
  E.~Komatsu {\it et al.}  [WMAP Collaboration],
  Astrophys.\ J.\ Suppl.\ {\bf 180}, 330 (2009).

\bibitem{Copeland:2006wr}
  E.~J.~Copeland, M.~Sami and S.~Tsujikawa,
  Int.\ J.\ Mod.\ Phys.\  D {\bf 15}, 1753 (2006)
  [arXiv:hep-th/0603057].
  
\bibitem{Bilic:2001cg}
  N.~Bilic, G.~B.~Tupper and R.~D.~Viollier,
  Phys.\ Lett.\  B {\bf 535}, 17 (2002)
  [arXiv:astro-ph/0111325].

\bibitem{Serra:2009yp}
  P.~Serra, A.~Cooray, D.~E.~Holz, A.~Melchiorri, S.~Pandolfi and D.~Sarkar,
  arXiv:0908.3186 [astro-ph.CO].

\bibitem{Weinberg72}
  S.~Weinberg,
  {\it Gravitation and Cosmology},
  (Wiley: New York, 1972) 685p.

\bibitem{Weinberg:1988cp}
  S.~Weinberg,
  Rev.\ Mod.\ Phys.\  {\bf 61}, 1 (1989).

\bibitem{Labun:2008qq}
  L.~Labun and J.~Rafelski,
  ``Dark Energy Content Of Nonlinear Electromagnetism,''
  arXiv:0811.4467 [hep-th].
and 
  ``QED Energy-Momentum-Trace in External Fields,''
  arXiv:0810.1323 [hep-ph].

\bibitem{Stone:1975bd}
  M.~Stone,
  Phys.\ Rev.\  D {\bf 14}, 3568 (1976).

\bibitem{Coleman:1977py}
  S.~R.~Coleman,
  Phys.\ Rev.\  D {\bf 15}, 2929 (1977)
  [Erratum-ibid.\  D {\bf 16}, 1248 (1977)].



\bibitem{Callan:1977pt}
  C.~G.~Callan and S.~R.~Coleman,
  Phys.\ Rev.\  D {\bf 16}, 1762 (1977).

\bibitem{Linde:1981zj}
  A.~D.~Linde,
  Nucl.\ Phys.\  B {\bf 216}, 421 (1983)
  [Erratum-ibid.\  B {\bf 223}, 544 (1983)].

\bibitem{BMTC}
B.~M\"uller and T.~Cohen, discussion remarks and private communications.
 

\bibitem{Einstein-lett}
  A. Einstein, in a letter to H.A.Lorentz, 15 November, 1919 Reported on page 2
    in {\it Einstein and the \ae ther}, L. Kostro (Apeiron: Montreal, 2000)].

\bibitem{Einstein-lect}
  A. Einstein, in inaugural lecture scheduled for May and given 27 October, 
  1920 at Reichs-Universit\"at zu Leiden
  [Einstein collected works (Springer: Berlin)].
  
\end{thebibliography}
\end{document}